\documentclass[aps,prd,preprintnumbers,nofootinbib,superscriptaddress,twocolumn]{revtex4-1}
\usepackage{natbib}
\usepackage{graphicx}
\usepackage{xcolor}

\usepackage{mathtools}
\usepackage{hyperref}
\usepackage{float}
\usepackage{soul}

\def\nat{Nature}

\begin{document}
\begin{flushright}
MI-TH-2025
\end{flushright}
\title{Constraints on MeV dark matter and primordial black holes: Inverse Compton signals at the SKA}

\author{Bhaskar Dutta}
\email{dutta@physics.tamu.edu}
\affiliation{Department of Physics and Astronomy, Mitchell Institute for Fundamental Physics and Astronomy, Texas A$\&$M University, College Station, TX 77843, USA}

\author{Arpan Kar}
\email{arpankar@hri.res.in}
\affiliation{Regional Centre for Accelerator-based Particle Physics, Harish-Chandra Research Institute, HBNI, Chhatnag Road, Jhunsi, Allahabad - 211 019, India}

\author{Louis E. Strigari}
\email{strigari@tamu.edu}
\affiliation{Department of Physics and Astronomy, Mitchell Institute for Fundamental Physics and Astronomy, Texas A$\&$M University, College Station, TX 77843, USA}

\begin{abstract}
We investigate the possibilities for probing MeV dark matter (DM) particles and primordial black holes (PBHs) (for masses $\sim 10^{15}$--$10^{17}$ g) 
at the upcoming radio telescope SKA, using photon signals from the Inverse Compton (IC) effect within a galactic halo. Pair-annihilation or 
decay of MeV DM particles (into $e^+ e^-$ pairs) or Hawking radiation from a population of PBHs generates mildly relativistic $e^{\pm}$ which can 
lead to radio signals through the IC scattering on low energy cosmic microwave background (CMB) photons. We study the ability of SKA to detect such 
signals coming from nearby ultra-faint dwarf galaxies Segue I and Ursa Major II as well as the globular cluster $\omega$-cen and the Coma cluster. 
We find that with $\sim 100$ hours of observation, the SKA improves the Planck constraints on the DM annihilation/decay rate and the PBH abundance for masses 
in the range $\sim 1$ to few tens of MeV and above $10^{15}$ to $10^{17}$ g, respectively.   
Importantly, the SKA limits are independent of the assumed magnetic fields within the galaxies. Previously allowed regions of diffusion parameters
of MeV electrons inside a dwarf galaxy that give rise to observable signals at the SKA are also excluded.
For objects like dwarf galaxies, predicted SKA constraints depend on both the DM and diffusion parameters. 
Independent observations in different frequency bands, e.g., radio and $\gamma$-ray frequencies, may break this degeneracy and thus enable one 
to constrain the combined parameter space of DM and diffusion. 
However, the constraints are independent of diffusion parameters for galaxy clusters such as Coma.
\end{abstract}


\maketitle

\section{Introduction}\label{sec:Introduction}

Dark matter (DM) has appeared as one of the major ingredients of the energy density of our universe \cite{Aghanim:2018eyx}, 
though its microscopic features are still unknown.
Weakly interacting, stable (on a cosmological time scale), MeV range beyond standard model (BSM) particles
are often proposed as viable candidates for DM \cite{Hooper:2008im, Dutra:2018gmv,Boddy:2015efa,Berlin:2018ztp,Dutta:2019fxn}.
Since no compelling evidence of usual weakly interacting massive particles
(WIMP), having masses in the GeV (or TeV) scale, has been found yet in 
any terrestrial DM search \cite{Aprile:2017iyp, PhysRevLett.118.251302, SHCHUTSKA2016656} or 
indirect search experiment \cite{Blanco_2019, Fermi-LAT:2016uux, Cavasonza:2016qem}, 
studies of sub-GeV DM particles have drawn recent interest~\cite{Battaglieri:2017aum}. 

Primordial black holes (PBHs) \cite{10.1093/mnras/152.1.75, 10.1093/mnras/168.2.399} 
are also used to explain the observed DM abundance of our universe and this idea has been explored 
through past several decades \cite{1975Natur.253..251C, Carr:2009jm}. 
In order to ensure that the lifetime of PBHs exceeds 
the age of the universe, 
their masses must be larger than $\sim 10^{15}$ g~\cite{Carr:2009jm}; 
for updated constraints see refs.~\cite{Poulter:2019ooo, Clark:2016nst, Carr:2016drx, PhysRevLett.125.101101}.  


Inside a DM halo, the annihilation and decay of MeV DM particles and the Hawking radiation from PBHs in the range $10^{15}$--$10^{17}$ g produce 
low energy $e^{\pm}$. Observing signals from these low energy $e^{\pm}$ in indirect detection experiments is a major challenge. 
The existing constraints from COMPTEL~\cite{Schoenfelder:2000bu} and INTEGRAL~\cite{Hermsen:2002qbl} exploit the photons produced from $e^{\pm}$. 
The Planck data on the cosmic microwave background (CMB) also provide stringent constraints
on DM particle masses as low as $\sim$ 1 MeV (in case of $e^+ e^-$ final state) \cite{Slatyer:2015jla, Slatyer:2016qyl} and 
for PBH masses in the range $10^{15}$ to $10^{17}$ g~\cite{Poulter:2019ooo}.  

In this paper, we describe the prospects for constraining the annihilation or decay rate of MeV DM particles and 
the abundance of PBHs in the search for photon signals generated through Inverse Compton (IC) scattering. 
The low energy $e^{\pm}$, emitted in the annihilation/decay of MeV DM particles 
or in the Hawking radiation from PBHs, produce those signals when 
scattering off the ambient photon bath.
While interactions with energetic CMB photons and other more energetic 
components of the bath such as infrared (IR) and starlight (SL)
lead to X-ray and soft $\gamma$-ray emissions \cite{Cirelli:2020bpc, Bartels:2017dpb},
scattering on the low-energy part of the CMB photon distribution can give rise to comparatively low frequency signals. 
Our aim is to study the possibility of detecting such low frequency  signals with
the upcoming radio telescope Square kilometer Array (SKA), which has been studied previously 
in the context of DM searches \cite{Colafrancesco:2014coa, Colafrancesco:2015ola, Caputo:2018ljp, Kar:2019cqo, Kar:2019mcq, Ghosh:2020ipv}.

To this point, galactic and extragalactic synchrotron fluxes have been studied  
using current and upcoming radio telescopes, which appear to be 
more effective in the case of annihilation or
decay of heavier DM particles \cite{Colafrancesco:2015ola, Kar:2019cqo, Kar:2019mcq, Ghosh:2020ipv, Cirelli:2016mrc, Regis:2017oet, Kar:2019hnj, Kar:2020coz}.
However, this method starts losing sensitivity (even for electron-positron dominated 
annihilations or decays) if masses of DM particles become 
$\lesssim$ a GeV.
This is also true in the case of Hawking radiation \cite{1974Natur.248...30H} 
(equivalent to the DM decay \cite{Clark:2016nst})
from PBHs with masses heavier than $10^{15}$ g.

We consider 100 hours of SKA observation of two local dwarf spheroidal (dSph) galaxies, namely, Segue I \cite{Natarajan:2015hma} and 
Ursa Major II \cite{Natarajan:2013dsa},
the globular cluster $\omega$-cen \cite{Brown:2019whs, Reynoso-Cordova:2019biv} 
and the Coma cluster \cite{Colafrancesco:2005ji} to determine the 
viable regions in the parameter spaces of MeV DM and PBH.
We compare our constraints with the existing CMB constraints obtained 
from Planck~\cite{Slatyer:2015jla, Slatyer:2016qyl, Poulter:2019ooo}, 
COMPTEL \cite{Essig:2013goa}, INTEGRAL \cite{Cirelli:2020bpc, Essig:2013goa, PhysRevD.101.123514}, and 
Voyager 1 \cite{Boudaud:2016mos, Boudaud:2018hqb}.
We exhibit how the radio survey of nearby dwarf galaxies through the SKA telescope can be
translated into the limits on the diffusion of sub-GeV electrons
inside those targets. 
Finally, we discuss the scope of the future observations of 
e-ASTROGAM \cite{DeAngelis:2017gra} in the context of MeV DM and PBH parameter spaces. 

The paper is organized as follows. In section \ref{sec:IC_flux}, we discuss IC fluxes from MeV DM and PBHs, 
in section \ref{sec:IC_calculation}, we discuss IC flux calculation, in section \ref{sec:Results} we present our results, and we conclude in section \ref{sec:Conclusion}.

\section{IC Fluxes from MeV DM and PBHs: radio signals for the SKA}\label{sec:IC_flux}

The upcoming radio telescope SKA is expected to play an important role in various fields of 
cosmology and astrophysics including the studies of DM \cite{Braun:2015B3, Power:2015G3, Colafrancesco:2015ola}.
The main advantage of the SKA lies not only in its 
capability of observing radio signal for a large frequency range (50 MHz - 50 GHz) which helps to constrain DM for a wide mass domain,
but also in its inter-continental baseline lengths which can well resolve the astrophysical 
foregrounds. Its large effective area helps it to achieve comparatively higher 
surface brightness sensitivity than any other existing radio telescope \cite{SKA}.

Studies of the particle nature of DM using SKA are based on its annihilation and decay 
properties \cite{Colafrancesco:2015ola, Ghosh:2020ipv}.
These annihilations or decays inside a dSph or galaxy cluster produce charged particles like electrons/positrons 
or neutral particles like photons. 
These particles may also originate from the Hawking radiation of PBHs, 
if it is assumed that a fraction of the DM abundance is made of them.
The $e^{\pm}$, upon interacting electromagnetically in the dSph or cluster medium, 
give rise to various types of photon fluxes such as Inverse Compton (IC), 
synchrotron etc \cite{Colafrancesco:2005ji, McDaniel:2017ppt}. Depending on the value of the
DM particle mass $m_{\chi}$ or PBH mass $M_{\rm PBH}$, peaks of these flux distributions may shift towards the higher 
or lower frequency side.

The synchrotron flux, produced from GeV (or TeV) DM particles in any dSph or galaxy cluster,
appears as a radio signal. Some previous works have studied this effect to constrain the 
GeV (or TeV) scale DM particles in the context of upcoming radio observations at the 
SKA \cite{Colafrancesco:2015ola, Colafrancesco:2014coa, Kar:2019mcq, Kar:2019cqo, Ghosh:2020ipv}. 
Nevertheless, it is difficult to detect the synchrotron flux in any radio telescope 
including the SKA if the DM particle mass reaches the MeV range. In this case, since the annihilation or decay 
spectra are lack of energetic $e^{\pm}$ which can 
interact with the magnetic field ($B$) present inside typical dSphs or galaxy clusters, the corresponding 
synchrotron emission is very weak in frequency.
Thus, constraining MeV DM particles, using the 
synchrotron radiation as a signal for the radio telescopes, is difficult.
This should also be true for PBHs which exist today and are able to emit $e^{\pm}$, i.e., 
with mass $M_{\rm PBH}$ in the range 
$10^{15} \rm{g}$ $\lesssim M_{\rm PBH} \lesssim$ $10^{17} \rm{g}$ \cite{Carr:2009jm}.

On the other hand, the photon flux generated in the IC scattering of electrons/positrons on 
CMB photons present inside a galaxy or galaxy cluster, 
is comparatively higher in frequency than the usual radio waves when one is looking for a $m_{\chi}$ 
in the GeV or TeV scale \cite{Colafrancesco:2005ji, Beck:2015rna, McDaniel:2017ppt} or a $M_{\rm PBH}$ which is much smaller 
than $10^{15}$ g (cannot exist today). 
In our current study show that if $m_{\chi}$ and $M_{\rm PBH}$ are, respectively, in the range 
$\sim 1$ to few tens of MeV and $\sim 10^{15}$ to $10^{17}$ g, the corresponding IC fluxes (or at least a part of their frequency distributions) 
can be observed at the SKA for DM particle and PBH parameter spaces that are consistent 
with existing experiments including the Planck and various radio observations. 
Interestingly, IC fluxes do not depend on the 
$B$-field of the parent galaxy or galaxy cluster, as strongly as in the case of synchrotron fluxes.

In addition to the IC flux that we consider, one can also try to look for the photon fluxes generated 
directly in the annihilation/decay of MeV DM particles or in the evaporation of PBHs with masses 
in the aforementioned range~\cite{Bartels:2017dpb, Boddy:2015efa, Carr:2009jm}

\section{IC flux calculation}\label{sec:IC_calculation}

\begin{figure*}[ht!]
\centering
  \includegraphics[width=8.9cm,height=7.7cm]{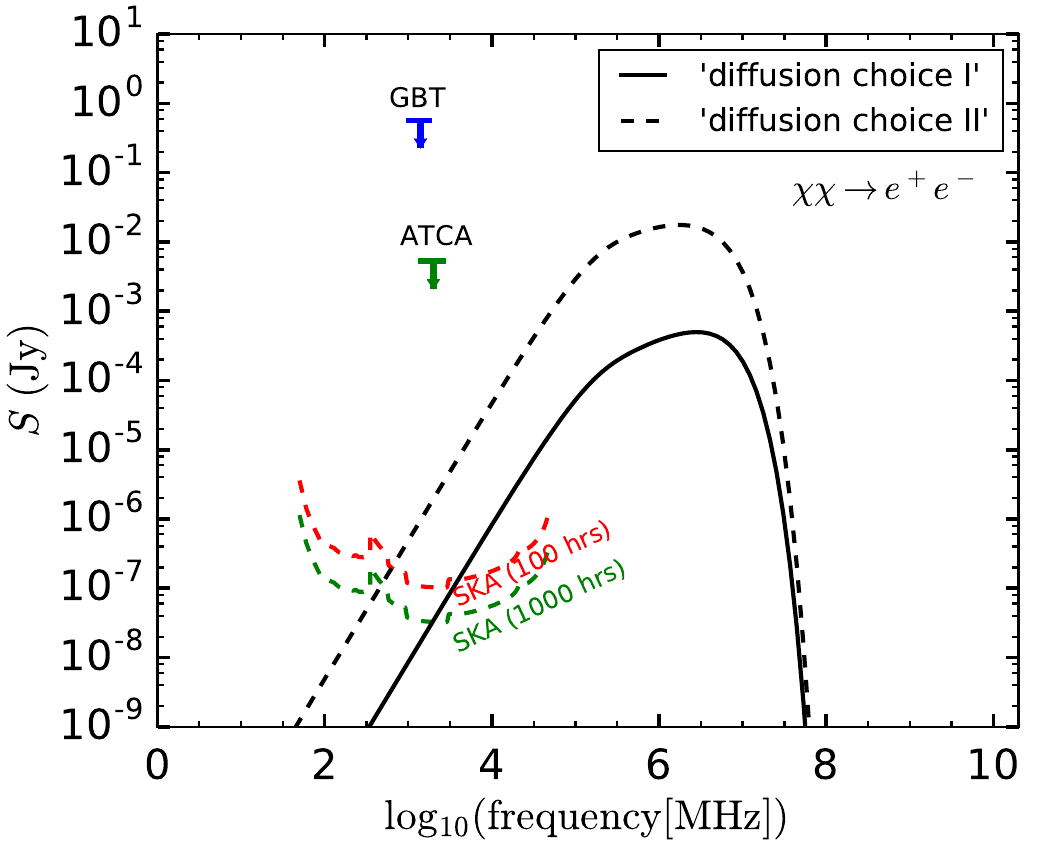}\hspace{0.06mm}
  \includegraphics[width=8.9cm,height=7.7cm]{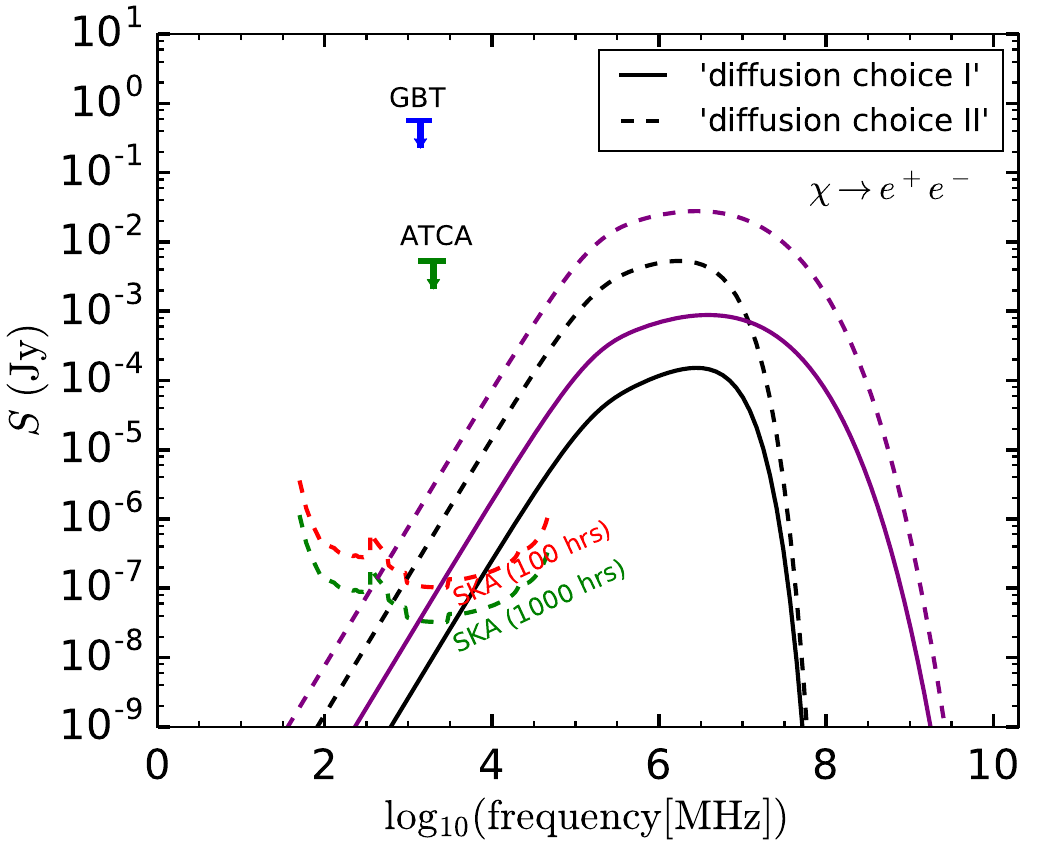}
\caption{{\it Left panel}: IC fluxes $S$ (black curves), generated in the annihilation of DM particles of
mass 2 MeV inside Segue I. The annihilation channel is assumed to be 
$\chi \chi \rightarrow e^+ e^-$ with an annihilation rate
$\langle \sigma v \rangle = 10^{-28} \rm{cm^3 s^{-1}}$. Parameter choices for diffusion are shown in the inset. 
The red and green dashed
lines represent the SKA sensitivities for 100 and 1000 hours, respectively \cite{Kar:2019mcq}. The blue and green arrows indicate the upper limits
on the radio flux obtained from GBT (1.4 GHz) \cite{Natarajan:2015hma, McDaniel:2017ppt} and ATCA ($\sim$ 2 GHz) \cite{Regis:2014koa} observations, respectively, 
for Segue I or dSph identical to Segue I. 
{\it Right panel}: Similar fluxes as shown in the left panel, but originating from the IC scattering of the $e^{\pm}$ produced in
a) the decay of 4 MeV DM ($\chi \rightarrow e^+ e^-$) with a decay width $\Gamma = 10^{-25} \rm{s^{-1}}$ (black curves); and 
b) the evaporation of PBHs with mass $M_{\rm PBH} = 10^{16}$ g and a population of 5\% of the total DM density of the dSph (purple curves).}
\label{IC_Flux}  
\end{figure*}

In order to estimate the IC flux, one first requires
the electron/positron energy spectrum originating from DM particles 
or PBHs. This can be done in terms of the
source function ($Q_e$) which, for annihilation/decay, is as \cite{Colafrancesco:2005ji, McDaniel:2017ppt, Regis:2017oet},

\begin{equation}
\begin{split}
Q^{\rm ann}_e(E,r) &= \frac{\langle \sigma v \rangle}{2 m_{\chi}^2} \rho^2(r) \frac{dN_e}{dE}, \\
Q^{\rm dec}_e(E,r) &= \frac{\Gamma}{m_{\chi}} \rho(r) \frac{dN_e}{dE},
\end{split}
\label{source_function}
\end{equation}
or for PBH evaporation is as \cite{Halzen:1995hu, PhysRevD.41.3052, Carr:2009jm},
\begin{equation}
Q^{\rm PBH}_e(E,r) = \frac{f_{\rm PBH}}{M_{\rm PBH}} \rho(r) \frac{d\dot{N}_e}{dE},
\label{source_function_PBH_1}
\end{equation}
with
\begin{equation}
\frac{d\dot{N}_e}{dE} = \frac{1}{2 \pi \hbar} \frac{\Gamma_e}{e^{\frac{E}{T_{\rm PBH}}} + 1}.
\label{source_function_PBH_2}
\end{equation}
Here $\langle \sigma v \rangle$ ($\Gamma$) is the annihilation (decay) rate of DM particles and
$\rho(r)$ is the DM density profile of the target, i.e., dSphs, or galaxy clusters, 
$\frac{dN_e}{dE}$ represents the electron (positron) energy distribution produced per
annihilation or decay. 
We assume that all the MeV DM particles annihilate or decay dominantly into $e^+ e^-$ final state.
The lifetime of the decaying DM is expected to exceed the age of the universe by several orders of magnitude \cite{Slatyer:2016qyl}.
The presence of $m_{\chi}^2$ or $m_{\chi}$ in the denominator of equation \ref{source_function} enhances the source function for MeV DM and thus causes
amplification in the signal.
In case of PBHs, $\frac{d\dot{N}_e}{dE}$ describes the electron/positron energy spectrum originating per unit time 
from the Hawking radiation of a single PBH. 
$\Gamma_e(E, M_{\rm PBH})$ is the absorption coefficient for spin-$\frac{1}{2}$ particles like electrons. An approximated analytic expression 
of $\Gamma_e(E, M_{\rm PBH})$ can be found in \cite{Halzen:1995hu, PhysRevD.41.3052}.
$T_{\rm PBH}$ represents the black hole temperature and is expressed as 
$T_{\rm PBH} = 1.06$ $\rm{GeV}$ $\times$ $\frac{10^{13} \rm g}{M_{\rm PBH}}$ \cite{Clark:2016nst}.
The $f_{\rm PBH}$ (in equation \ref{source_function_PBH_1}) denotes the PBH fraction of the total DM density $\rho(r)$.
All the PBHs are assumed to have the same mass $M_{\rm PBH}$ and 
their lifetimes are taken to be larger than the age of the universe. 
The latter assumption is satisfied by the values of $M_{\rm PBH}$ 
considered for the current work.

The $e^{\pm}$, after being emitted, 
propagate through the galactic or cluster medium and give rise to a equilibrium density
$\frac{dn_e}{dE}(E, r)$ which one can obtain by solving the transport
equation \cite{Colafrancesco:2005ji, McDaniel:2017ppt, Colafrancesco:2006he},

\begin{equation}
D(E) \nabla^2 \left(\frac{dn_e}{dE}\right) +
\frac{\partial}{\partial E}\left(b(E) \frac{dn_e}{dE}\right) +
Q_e = 0,
\label{transeport_equation}
\end{equation}
with $D(E)$ and $b(E)$ as the diffusion and energy loss terms, respectively. 
The IC flux $S(\nu)$, as a function of the observation frequency $\nu$, is acquired by folding this $\frac{dn_e}{dE}$ with
the IC power spectrum $P_{\rm IC}(\nu,E)$ and integrating over the emission region ($\Omega$) of 
the target \cite{Colafrancesco:2005ji, McDaniel:2017ppt},

\begin{equation}
S(\nu) = \frac{1}{4\pi} \int d\Omega \int_{\rm los} dl \left(2 \int dE\frac{dn_e}{dE} P_{\rm IC}\right),
\label{S_IC}
\end{equation}
where $l$ is the line-of-sight (los) co-ordinate.
The IC power spectrum $P_{\rm IC}(\nu,E)$ can be obtained by multiplying the IC scattering cross section 
($\sigma_{\rm IC}(\nu,\epsilon,E)$) with the CMB photon number density ($n(\epsilon)$)
and integrating over the appropriate range of energy $\epsilon$,\footnote{CMB photon energy $\epsilon$ which produces the IC flux at $\nu = 10$ GHz for $E = 2$ MeV, 
lies typically in the range 
$\sim 7\times10^{-13}$ -- $\sim 4\times10^{-11}$ MeV (i.e. $\sim 0.2$ -- $\sim 10$ GHz, if expressed in terms of the CMB photon frequency).}
(see \cite{Colafrancesco:2005ji, McDaniel:2017ppt, Cirelli:2010xx, Cirelli:2020bpc}).

In case of MeV electrons, we have parameterised the diffusion term in equation \ref{transeport_equation} as \cite{Natarajan:2015hma, McDaniel:2017ppt, Natarajan:2013dsa},
\begin{equation}
D(E) = D_0 \left(\frac{E}{E_0}\right)^{\gamma},  
\label{D_E}
\end{equation}
where $D_0$ and $\gamma$ are, respectively, the diffusion coefficient and diffusion index and $E_0 = 1$ GeV. 
Values of these parameters inside any galactic system 
are not very well constrained for sub-GeV electrons \cite{Bartels:2017dpb}.
We have used some illustrative values of them, as well as scanned the possible range of parameter space. 
The term $b(E)$ in equation \ref{transeport_equation} describes the energy loss processes of the $e^{\pm}$ 
due to various types of electromagnetic effects such as
IC scattering, synchrotron radiation, coulomb interaction, and bremsstrahlung.
Readers are referred to \cite{Colafrancesco:2005ji, McDaniel:2017ppt, Colafrancesco:2006he, Kar:2020coz} 
for the detailed parameterisation of $b(E)$ inside various targets like the galaxy clusters and dSphs. 
Note that the IC and synchrotron losses are 
proportional to the square of the electron energy ($E$) and hence are suppressed for MeV electrons 
in comparison to other loss processes.
Due to the proximity of the local dSphs like the Segue I, the magnetic field strength and parameterisation of diffusion in them 
are influenced by the choices used for 
the Milky Way galaxy \cite{Regis:2014koa, Jeltema:2008ax, Natarajan:2015hma}.

In Figure \ref{IC_Flux}, IC fluxes ($S(\nu)$) from the nearby dSph Segue I are shown for 
annihilating DM mass $m_{\chi} = 2$ MeV (left panel; black curves), decaying DM mass $m_{\chi} = 4$ MeV 
(right panel; black curves) and evaporating PBH mass $M_{\rm PBH} = 10^{16}$ g (right panel; purple curves). 
The DM distribution $\rho(r)$ for this dSph is assumed to follow a Einasto profile \cite{1965TrAlm...5...87E} 
with parameters similar to the ones found in \cite{Natarajan:2015hma, McDaniel:2017ppt, Aleksic:2011jx}. 
The annihilation (decay) final state is assumed to be $e^+ e^-$ with a rate
$\langle \sigma v \rangle = 10^{-28} \rm{cm^3 s^{-1}}$ 
($\Gamma = 10^{-25} \rm{s^{-1}}$). In case of PBHs, it is assumed that these objects cover 5\% of the total DM density of 
the dSph and all of them have the same mass $M_{\rm PBH}$.
Fluxes are estimated for two 
illustrative choices of $D_0$ and $\gamma$, namely, 
(I) $D_0 = 2.3 \times 10^{28} \rm cm^2 s^{-1}$; $\gamma = 0.46$ (solid curves) 
and (II) $D_0 = 3 \times 10^{27} \rm cm^2 s^{-1}$; $\gamma = 0.7$ (dashed curves) \cite{Natarajan:2015hma, McDaniel:2017ppt}.
We refer them as `diffusion choice I' and `diffusion choice II', respectively.
These values are often used in the context of the Milky way galaxy \cite{Buch:2015iya, Lavalle:2014kca, Boudaud:2016mos}. 
The higher $D_0$ reduces the equilibrium electron/positron density $\left(\frac{dn_e}{dE}\right)$ in equation \ref{transeport_equation} 
and hence the flux in equation \ref{S_IC} (see reference \cite{Kar:2019cqo}).
Expected SKA sensitivities in the frequency range 50 MHz - 50 GHz 
for 100 and 1000 hours of observations are shown by the red and green dashed curves, respectively \cite{Kar:2019mcq}.
These sensitivities have been calculated using the documents provided in the SKA website \cite{SKA}. 
See \cite{Kar:2019cqo} for details of the analysis.
Along with these, upper limits on the 
radio signal from the observations of Green Bank Telescope (GBT) \cite{Natarajan:2015hma, McDaniel:2017ppt} and 
Australia Telescope Compact Array (ATCA) \cite{Regis:2014koa} towards Segue I or dSph similar
to Segue I are also shown in both panels of the same figure by the blue and green arrows, respectively. 
It can be seen that, with parameter choices mentioned above, it is possible for the MeV DM or PBH induced IC fluxes to overcome the SKA threshold
(mainly in the high frequency range, i.e., for $\nu \gtrsim 1$ GHz),
by maintaining constraints from existing radio experiments.

One important point regarding the IC fluxes shown in Figure \ref{IC_Flux} is that, unlike synchrotron fluxes, 
they are weakly-dependent on the magnetic field ($B$). The main reason behind this is the absence of the $B$-field in 
the IC power spectrum $P_{\rm IC}$ \cite{Colafrancesco:2005ji, McDaniel:2017ppt, Cirelli:2010xx}, 
appearing in equation \ref{S_IC}. There can be a $B$ dependence in the 
$\frac{dn_e}{dE}$ itself, through the energy loss term $b(E)$ (see equation \ref{transeport_equation}) 
pertaining to the synchrotron effect 
(which goes as $B^2 E^2$) \cite{Colafrancesco:2005ji, McDaniel:2017ppt, Colafrancesco:2006he}. 
However, as pointed out earlier, this process is usually suppressed in case of
MeV electrons for typical values of $B$ present inside the 
galaxy clusters and dSphs \cite{Regis:2014koa, Colafrancesco:2005ji}. We have checked that the fluxes in Figure \ref{IC_Flux} vary at most 
by $\sim 6\%$ (at $\nu = 10$ GHz) for different values of $B$ 
in the range 0 to 10 $\mu G$. 
Due to this very small variation, we can say that
the results presented in this figure and hereafter are almost 
independent of the magnetic field $B$.

As a target for the MeV DM or PBH search in the context of SKA, we have mainly used the nearby
ultra-faint dSphs such as Segue I and Ursa Major II \cite{Natarajan:2015hma, McDaniel:2017ppt, Natarajan:2013dsa} in our work. 
These galaxies are appropriate for studying DM induced signals due to their 
low star formation rates which minimise the contribution of astrophysical processes. 
Their high DM contents (as inferred from their high mass-to-light ratios) and close proximity 
are of additional advantages. 
For comparison, we also present the results for the globular cluster
$\omega$-cen \cite{Kar:2020coz, Brown:2019whs, Reynoso-Cordova:2019biv} and a galaxy cluster identical to the Coma cluster \cite{Colafrancesco:2005ji}. 
The $\omega$-cen is almost ten times closer than any nearby dSph and may have DM density as high as 
any compact dwarf \cite{Brown:2019whs}. These make it useful for studies of DM \cite{Kar:2020coz}. 
The Coma cluster, on the other hand, also contains a significant amount of DM. As described in \cite{Colafrancesco:2005ji, McDaniel:2017ppt}, 
due to its large radius, the solution to the equation \ref{transeport_equation} is independent of diffusion inside this target.

\section{Results}\label{sec:Results}

\begin{figure*}[ht!]
\centering
  \includegraphics[width=8.9cm,height=8cm]{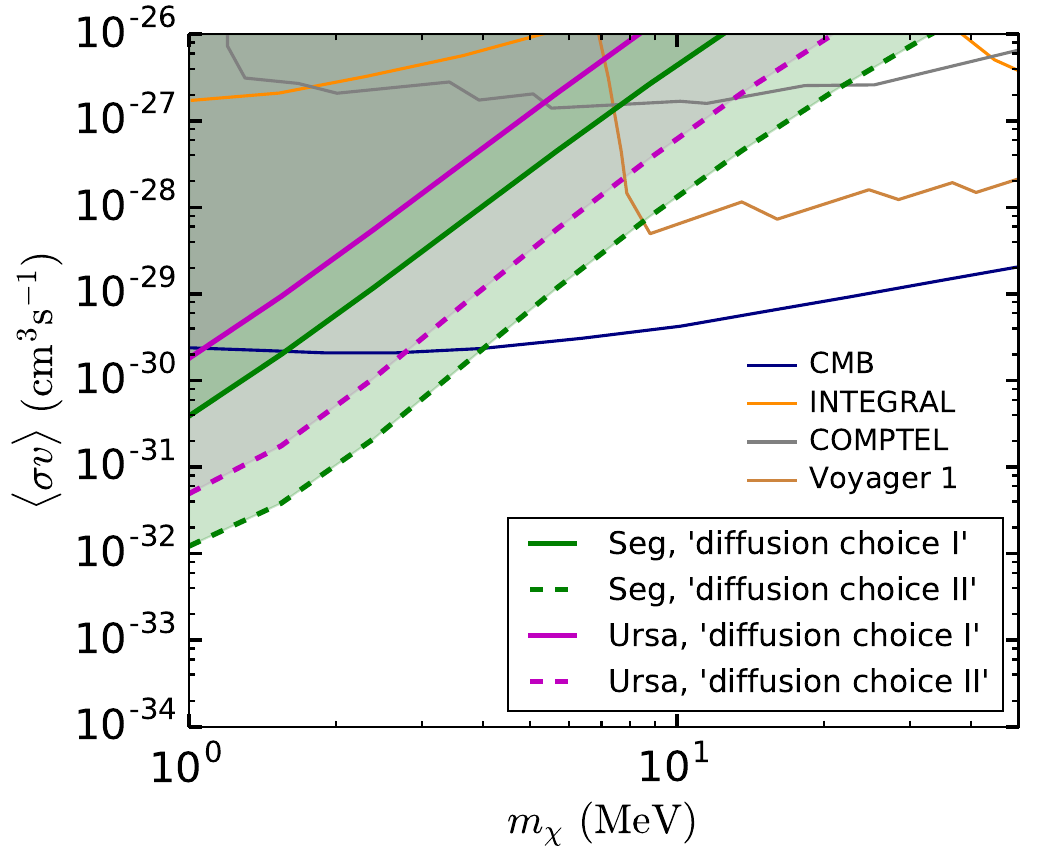}
  \includegraphics[width=8.9cm,height=8cm]{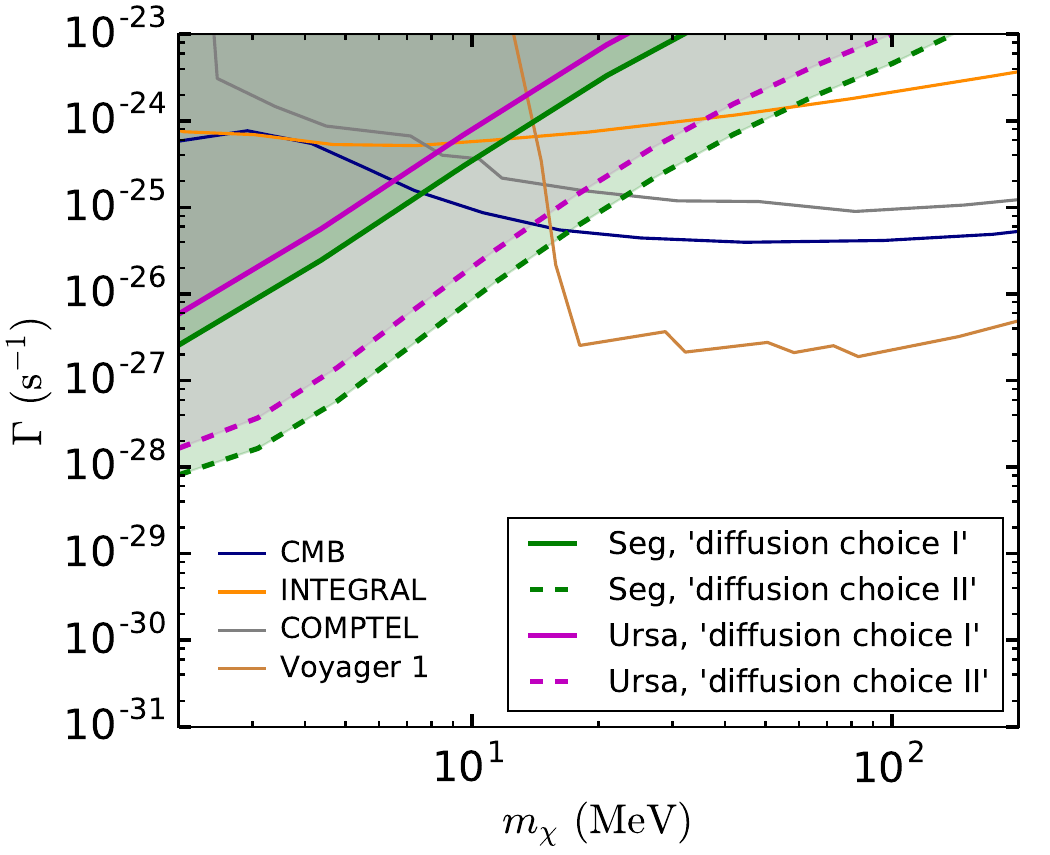}
  \includegraphics[width=8.9cm,height=8cm]{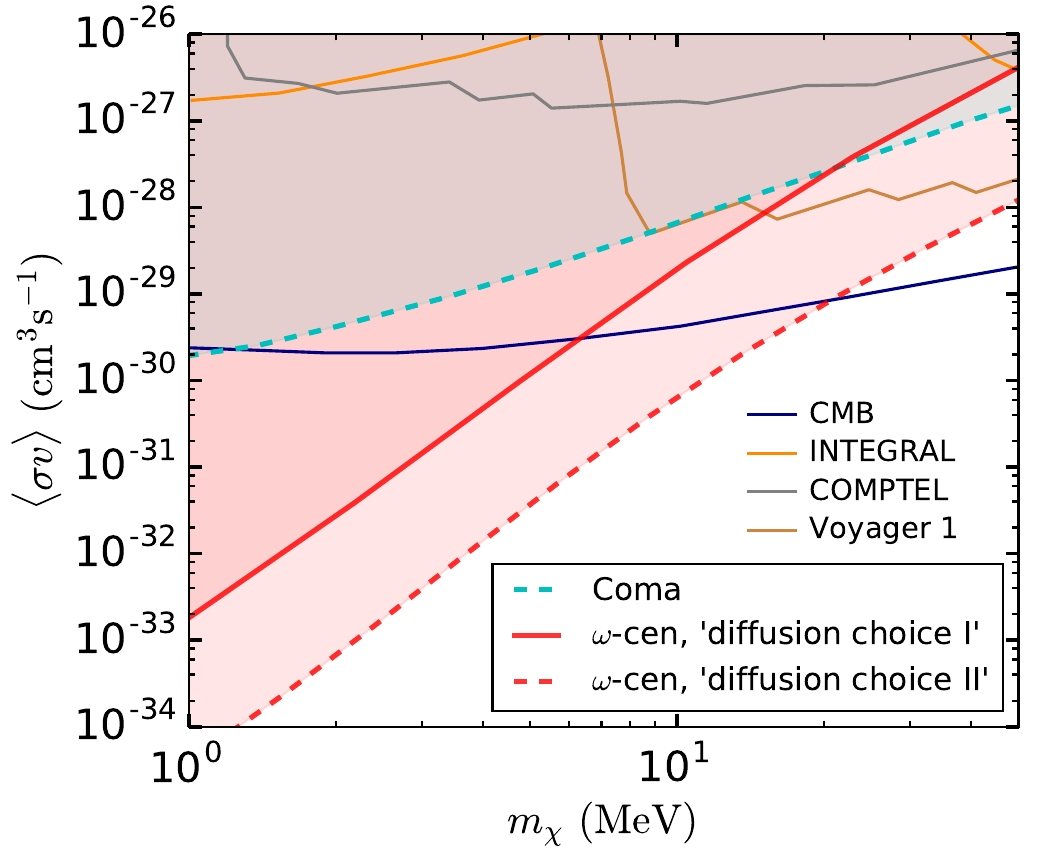}
  \includegraphics[width=8.9cm,height=8cm]{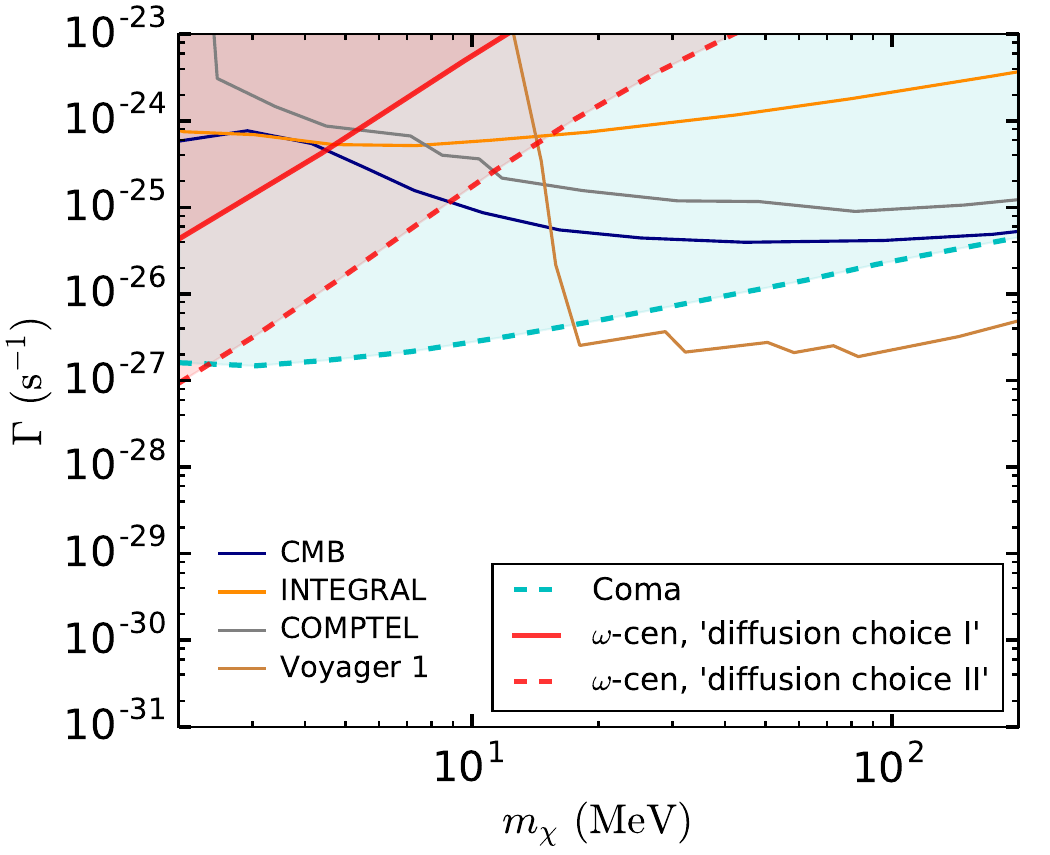}
\caption{The thick solid and dashed lines (with shaded regions) represent the SKA threshold limits in the $\langle \sigma v \rangle - m_{\chi}$ (left
column) and $\Gamma - m_{\chi}$ (right column) planes for detecting (in 100 hours observation) the DM induced IC fluxes produced in the $e^+ e^-$ channel. The limits
are shown for different targets: Segue I (in green), Ursa Major II (in magenta), 
$\omega$-cen (in red) and Coma cluster (in cyan). Parameter choices for 
diffusion are shown in the inset. In case of Coma, the
limits are independent of $D_0$. 
Constraints coming from CMB data \cite{Slatyer:2015jla, Slatyer:2016qyl}, Voyager 1 \cite{Boudaud:2016mos}, 
INTEGRAL \cite{Cirelli:2020bpc, Essig:2013goa} and COMPTEL \cite{Essig:2013goa} are 
shown by various thin solid lines.}
\label{DM_limit}  
\end{figure*}

\begin{figure*}[ht!]
\centering
  \includegraphics[width=8.9cm,height=8cm]{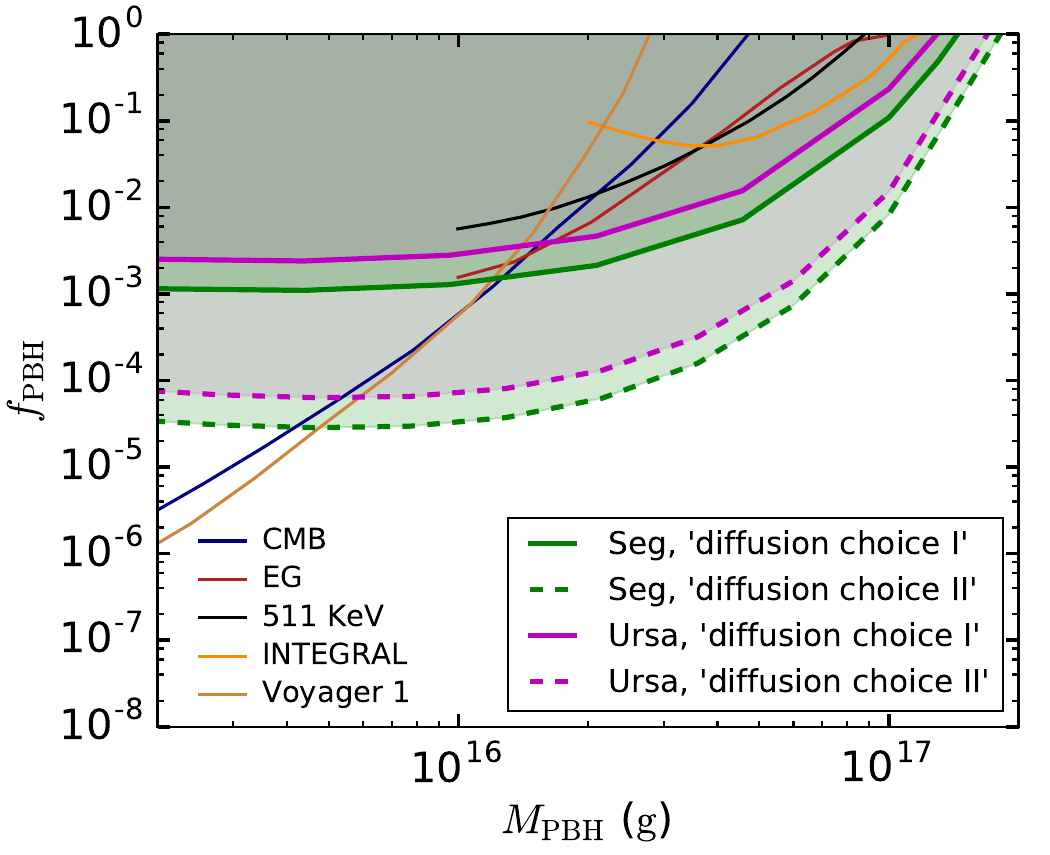}
  \includegraphics[width=8.9cm,height=8cm]{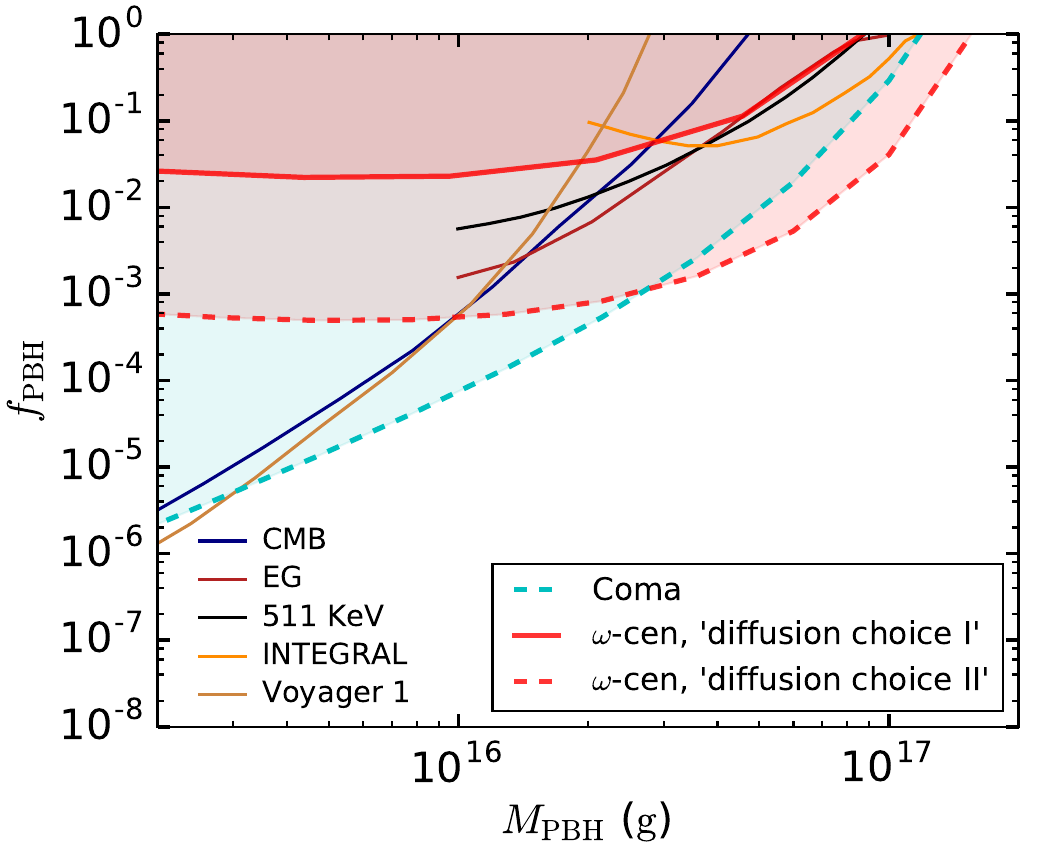}
\caption{SKA 100 hours threshold limits (shown by the thick solid and dashed curves with shaded regions) in the $f_{\rm PBH} - M_{\rm PBH}$ plane for observing 
the IC fluxes which are produced by the IC scattering of the $e^{\pm}$ emitted in PBH evaporation. The limits
are presented for different targets: Segue I and Ursa Major II (left panel), 
$\omega$-cen and Coma cluster (right panel). 
Parameter choices for diffusion are in the inset. 
The limit in case of Coma is independent of $D_0$. 
Constraints from  CMB \cite{Poulter:2019ooo}, extra-galactic (EG) $\gamma$-ray emission \cite{Ballesteros:2019exr}, 
511 keV line emission \cite{Laha:2019ssq}, INTEGRAL \cite{PhysRevD.101.123514} and Voyager 1 \cite{Boudaud:2018hqb} 
are shown by thin solid lines.}
\label{PBH_limit}  
\end{figure*}

In Figure \ref{DM_limit}, thick solid and dashed curves show the threshold limits in 
the $\langle \sigma v \rangle - m_{\chi}$ (left column) 
and $\Gamma - m_{\chi}$ (right column) planes for observing any IC effect induced radio 
signal at the SKA with 100 hours of observation time.
The DM annihilation and decay scenarios are $\chi \chi \rightarrow e^+ e^-$ and $\chi \rightarrow e^+ e^-$, respectively.
The limits are shown for Segue I (green lines) and Ursa Major II (magenta lines) together with 
$\omega$-cen (red lines) and Coma cluster (cyan lines). 
These limits are calculated following the analysis presented in \cite{Kar:2019cqo}.
As mentioned regarding the discussion of Figure \ref{IC_Flux}, the DM density distribution for Segue I is described by a 
Einasto profile with profile information same as in 
\cite{Natarajan:2015hma, McDaniel:2017ppt, Aleksic:2011jx}.
For the other dSph Ursa major II and the globular cluster $\omega$-cen, we assume the Navarro-Frenk-White (NFW) \cite{Navarro:1995iw} 
DM profile following the parameterisations in references 
\cite{Natarajan:2013dsa} and \cite{Brown:2019whs}, respectively. In case of Coma we take a N04 profile \cite{Navarro:2003ew} to represent the total DM distribution
and along with this, add substructure contributions for the annihilation scenario (see \cite{Colafrancesco:2005ji} for details).
Choices of diffusion parameters ($D_0$ and $\gamma$) for the dSph galaxies and the $\omega$-cen are: 
$D_0 = 2.3 \times 10^{28} \rm cm^2 s^{-1}$; $\gamma = 0.46$ (`diffusion choice I' -- thick solid curves) 
and $D_0 = 3 \times 10^{27} \rm cm^2 s^{-1}$; $\gamma = 0.7$ (`diffusion choice II' -- thick dashed curves) 
\cite{Buch:2015iya, Boudaud:2016mos, Natarajan:2015hma, McDaniel:2017ppt, Natarajan:2013dsa}.  
The diffusion time-scale of MeV electrons in a big system like the Coma cluster is typically larger than their energy 
loss time-scale \cite{Colafrancesco:2005ji, McDaniel:2017ppt}.
As a result, the IC fluxes from Coma and hence the 
corresponding SKA limits (shown in this figure and the next) are independent of the diffusion.
In addition to the SKA predictions, in each panel of Figure \ref{DM_limit}, 
constraints obtained using 
Planck's CMB data \cite{Slatyer:2015jla, Slatyer:2016qyl}, 
diffuse X-ray and $\gamma$-ray data from INTEGRAL \cite{Cirelli:2020bpc, Essig:2013goa} and COMPTEL \cite{Essig:2013goa} 
and $e^{\pm}$ data from Voyager 1 \cite{Boudaud:2016mos} are 
indicated by various thin solid lines with different colors.

For dSphs and $\omega$-cen, the SKA limits shown in Figure \ref{DM_limit} depend on the choices of diffusion parameters but not on the magnetic field $B$ (see the earlier discussions). 
For Coma they also are independent of the diffusion.
In each case, the regions above the SKA limits and below the Planck and other existing constraints are 
the parameter spaces that are available for the future radio telescope SKA. 
In case of both annihilation and decay, these regions extend up to DM mass $m_{\chi} \simeq 20$ MeV. 

Figure \ref{PBH_limit} demonstrates the SKA threshold limits (thick solid and dashed lines) on the 
PBH fraction $f_{\rm PBH}$ as a function of the PBH mass ($M_{\rm PBH}$). 
Side-by-side, constraints estimated using  
CMB \cite{Poulter:2019ooo}, extra-galactic (EG) $\gamma$-ray flux \cite{Ballesteros:2019exr}, 
511 KeV line \cite{Laha:2019ssq}, 
Voyager 1 \cite{Boudaud:2018hqb} and INTEGRAL \cite{PhysRevD.101.123514} data are shown by different thin solid lines. 
The SKA limits for the dSphs Segue I and Ursa Major II are presented in the left panel of the figure, 
while the right panel depicts the limits 
for $\omega$-cen and Coma cluster.
Choices of all the astrophysical parameters are exactly the same as in Figure \ref{DM_limit}.
The PBH parameter space that is consistent with existing data and can be constrained through the SKA is stretched over the PBH mass range $\sim 3 \times 10^{15}$ -- $\sim 10^{17}$ g. 
Note that PBHs with mass $M_{\rm PBH} < 10^{15}$ g do not exist in today's galactic halos, whereas 
for $M_{\rm PBH} > 10^{17}$ g, it is difficult to produce $e^{\pm}$ (which in turn generate the IC flux) in the Hawking radiation \cite{Carr:2009jm}.

\begin{figure*}[ht!]
\centering
  \includegraphics[width=8.9cm,height=8cm]{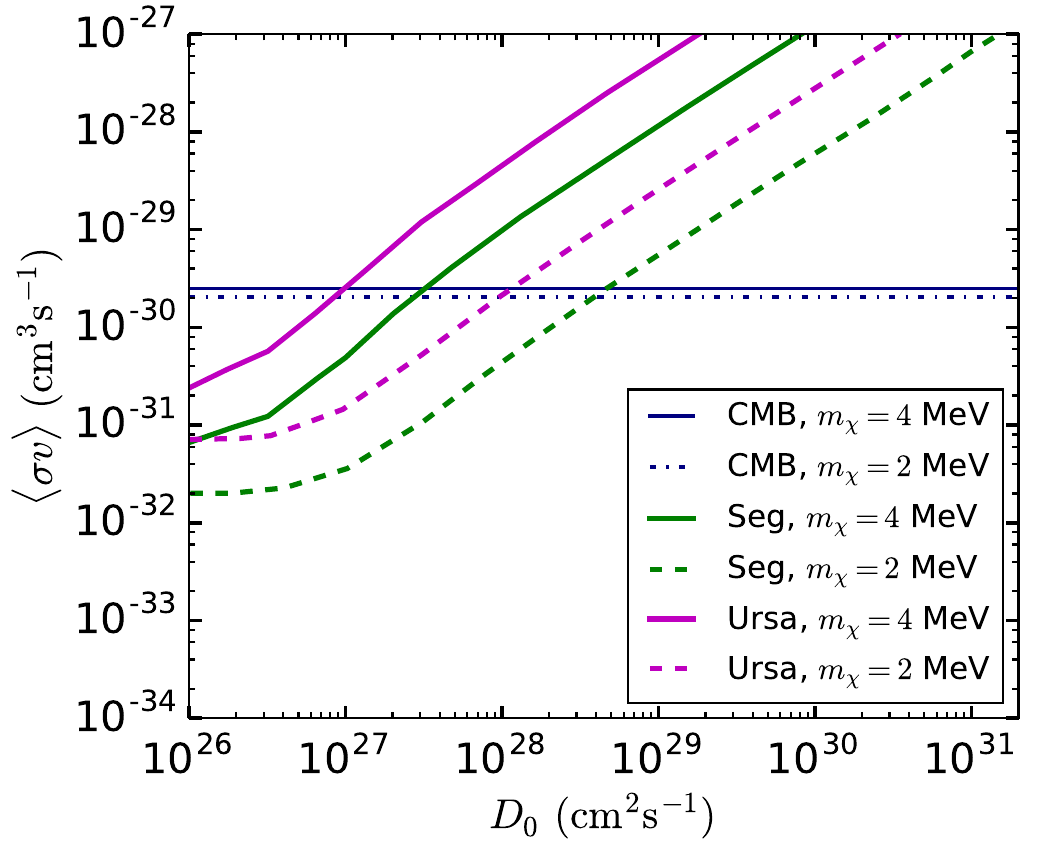}
  \includegraphics[width=8.9cm,height=8cm]{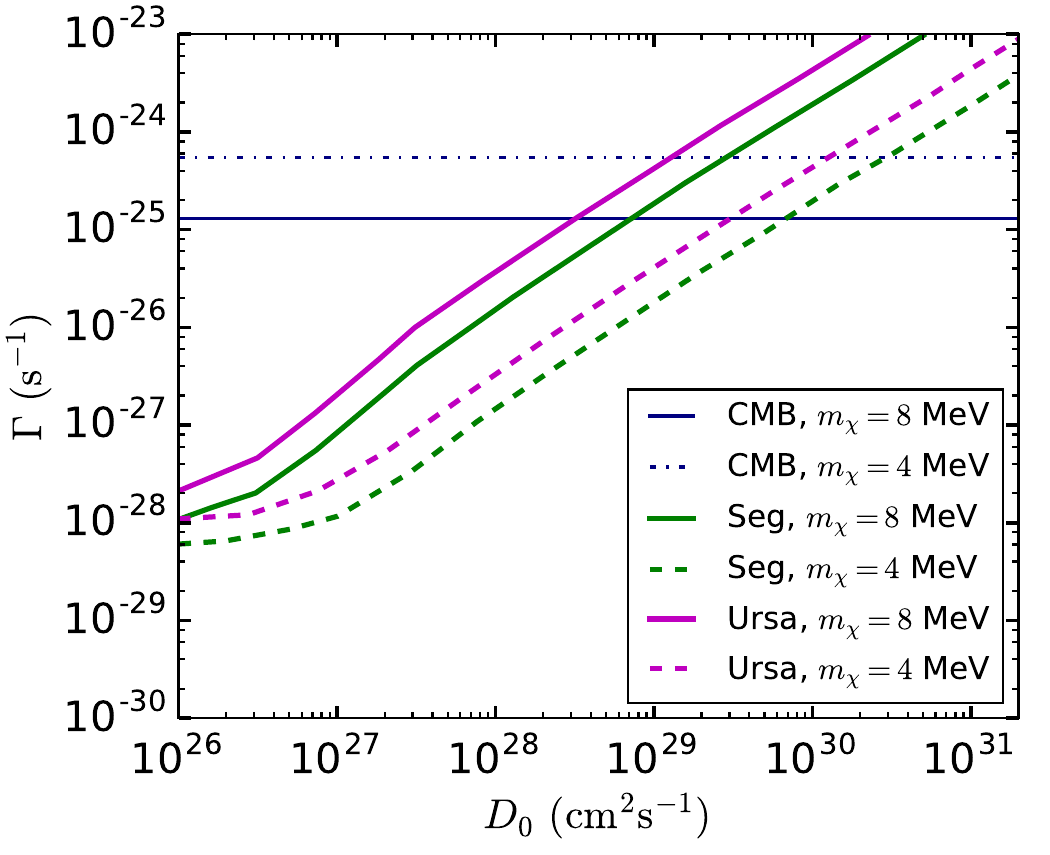}
\caption{Limits in the $\langle \sigma v \rangle - D_0$ (left panel) and $\Gamma - D_0$ (right panel) planes to observe the DM induced IC fluxes at SKA (100 hours) 
from Segue I (green lines) and Ursa Major II (magenta lines) for different DM masses shown in the inset. 
The final state of annihilation and decay is $e^+ e^-$. The limits are shown for $\gamma$ = 0.7 \cite{Natarajan:2015hma, Natarajan:2013dsa}. 
Corresponding CMB upper limits at those DM masses are indicated by the solid and dashed-dotted dark blue lines \cite{Slatyer:2015jla, Slatyer:2016qyl}.}
\label{DM_limit_D0}  
\end{figure*}

\begin{figure}[hbt!]
\centering
  \includegraphics[width=8.6cm,height=7.7cm]{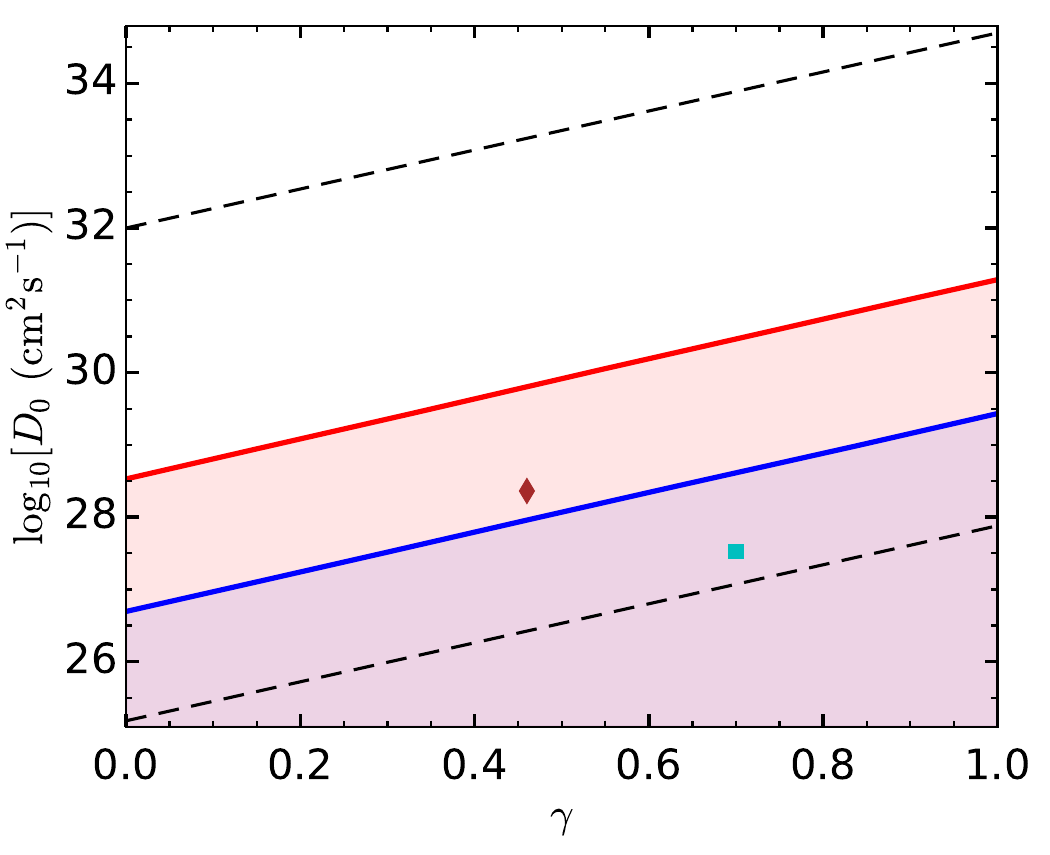}
\caption{The blue (red) line represents the SKA 100 hours
threshold limit in the $D_0 - \gamma$ plane for observing the IC flux produced by DM of mass $m_{\chi}$ = 2 (4) MeV
annihilating (decaying) into the $e^+ e^-$ channel from Segue 1. The corresponding $\langle \sigma v \rangle$ ($\Gamma$) is consistent with the upper limit obtained 
by Planck (see Figure \ref{DM_limit_D0}). The shaded region below each line
indicates the parameter space which can be probed or ruled
out by the SKA. 
Two illustrative values of $D_0$ and $\gamma$, used in Figures 
\ref{IC_Flux}, \ref{DM_limit} and \ref{PBH_limit}, are marked with brown and cyan points.
The upper dashed line shows the maximum $D_0$ (at $E = 2$ MeV) 
allowed in a typical dwarf galaxy,
while the lower dashed line depicts the limit on $D_0$ (at the same $E$), below which the IC flux from the galaxy is not sensitive to the diffusion (see the text for details).}
\label{D0_index}  
\end{figure}

The SKA limits in Figures \ref{DM_limit} and \ref{PBH_limit} assume
some illustrative values of the diffusion coefficient $D_0$ and index $\gamma$.
However, the lack of observational data makes it difficult to constrain these parameters in the MeV energy scale, for the galaxies including our own 
\cite{Bartels:2017dpb}.
Keeping this fact in mind, we have varied those parameters over possible ranges and 
shown the allowed regions which can give rise to detectable signals at the SKA.

In Figure \ref{DM_limit_D0}, SKA threshold limits on DM annihilation rate $\langle \sigma v \rangle$ (left panel)
and decay rate $\Gamma$ (right panel) as a function of $D_0$ have been 
presented for the dSphs Segue I (green lines) and Ursa Major II (magenta lines), 
assuming $\gamma = 0.7$ \cite{Natarajan:2015hma, McDaniel:2017ppt, Natarajan:2013dsa}. 
These limits are for DM masses $m_{\chi} = 2$ and 4 MeV 
(in case of annihilation) and $m_{\chi} = 4$ and 8 MeV (in case of decay). 
Simultaneously, we have also shown the corresponding Planck's CMB constraints at those masses.
The intersections of the SKA limits and the Planck's constraints indicate the ranges
of $D_0$ (with $\gamma = 0.7$) which can be probed at the SKA for the dSphs under consideration.   
Detections of the IC effect induced radio signals for $D_0$ beyond these ranges would require such annihilation or decay rates 
that are already ruled out by the Planck. Similar types of plots can be obtained for the PBH fraction ($f_{\rm PBH}$) too.

Using some independent information on the MeV DM mass and the corresponding annihilation or decay rate, 
we identify the viable region in the $D_0 - \gamma$ plane, 
as has been shown in Figure \ref{D0_index}. Here, the blue and red lines indicate the SKA limits in this plane 
for detecting the IC fluxes originating, respectively, from the annihilation and decay of DM particles into $e^+ e^-$ final state. 
These limits are estimated for the dSph Segue I.
The DM masses are assumed to be $m_{\chi} = 2$ MeV (for annihilation) and $m_{\chi} = 4$ MeV (for decay). 
The corresponding $\langle \sigma v \rangle$ and $\Gamma$ are kept fixed at the upper limits obtained from 
Planck's CMB observation. 
The index $\gamma$ has been scanned over the range $0 \leq \gamma \leq 1$ \cite{Bartels:2017dpb, Lavalle:2014kca, Jeltema:2008ax}.
For electron energies in the sub-GeV scale, the diffusion (in equation \ref{D_E}) gets stronger for lower $\gamma$ and 
therefore suppresses the IC flux (see equations \ref{transeport_equation} and \ref{S_IC}).
As a result, the limits in the $D_0 - \gamma$ plane become weaker for the smaller values of $\gamma$. 
The shaded region under each line in this figure represents the parameter space to be constrained by the SKA. 
For any $\gamma$, the values of $D_0$ above the blue (red) line will make the radio signals detectable at the SKA for $\langle \sigma v \rangle$ ($\Gamma$) larger than 
the Planck limit (see Figure \ref{DM_limit_D0}). 
Note that, unlike in the case of synchrotron radiation \cite{Natarajan:2015hma, Natarajan:2013dsa}, 
radio limits presented here depend very feebly on the magnetic field $B$. 
Besides the SKA limits, we have shown another two possible bounds on $D_0$ in Figure \ref{D0_index}. 
The upper black dashed line indicates the maximum $D_0$ (at electron energy $E = 2$ MeV) allowed in a dSph, having a length-scale $\approx 1$ kpc.
For $D_0$ above this line, the MeV electrons can have a diffusion velocity 
greater than the speed of light (c) (see the arguments of \cite{Regis:2014koa}).
The lower black dashed line, on the other hand, shows the limit on $D_0$ (at $E = 2$ MeV), below which the diffusion time-scale of the $e^{\pm}$
in the dSph becomes larger than their energy loss time-scale and the corresponding IC fluxes are not sensitive to the diffusion anymore 
(see the discussion in \cite{Colafrancesco:2005ji}).
One can repeat the whole analysis, described here, for the other dSph or $\omega$-cen and simultaneously use the information on PBH in place of MeV DM. 
The measurements associated with the Segue I, Ursa Major II and the globular cluster $\omega$-cen depend on $D_0$. 
However, the Coma cluster observations are independent of $D_0$. It is also possible to combine these measurements to determine $D_0$ and $\gamma$.

Apart from the SKA radio telescope project, various space based MeV range $\gamma$-ray experiments 
such as e-ASTROGAM \cite{DeAngelis:2017gra}, AMEGO \cite{McEnery:2019tcm}, GRAMS \cite{Aramaki:2019bpi}, have been recently  
proposed and are expected to start their operations in the forthcoming years. These experiments will also play important roles 
in probing DM and PBHs in the mass domains considered in this work. 
For example, e-ASTROGAM, due its comparatively higher sensitivity \cite{DeAngelis:2017gra}, 
can constrain the DM annihilation rate and decay width (for $e^+ e^-$ final state) up to the 
values $\langle \sigma v \rangle \simeq 10^{-30} \rm cm^3 s^{-1}$ and $\Gamma \simeq 5 \times 10^{-27} \rm s^{-1}$ respectively, 
for a DM mass ($m_{\chi}$) $\simeq 2$ MeV \cite{Bartels:2017dpb}. 
These constraints are stronger by a few orders of magnitude in comparison to the corresponding limits coming from existing MeV $\gamma$-ray 
experiments like COMPTEL and INTEGRAL~\cite{Essig:2013goa}. 
A similar conclusion also holds for PBHs with masses that lie between 
$10^{15}$ and $10^{17}$ g \cite{DeAngelis:2017gra}. 
The future experiment AMEGO is predicted to give more or less same results as e-ASTROGAM \cite{Bartels:2017dpb, Ballesteros:2019exr}. 
The GRAMS satellite mission, on the other hand, can have slightly better sensitivity 
(with a sufficient observation time) in measuring 
MeV photon signals \cite{Aramaki:2019bpi} 
and thus may provide even more stringent constraints on both the DM and PBH parameter spaces.
By comparing the aforementioned e-ASTROGAM estimations with the SKA limits from Figures \ref{DM_limit} and \ref{PBH_limit}, 
it can be seen that the SKA, in some cases, can have better constraints for $\langle \sigma v \rangle$ 
depending on the diffusion in the system, 
but independent of diffusion for $\Gamma$ and $f_{\rm PBH}$ if one looks for a cluster scale object like Coma.

\section{Conclusion}\label{sec:Conclusion}

Investigating MeV scale signals for dark matter and PBHs is well motivated. 
For example, MeV scale dark matter searches are underway at 
ongoing and upcoming direct detection~\cite{Barak:2020fql, Essig:2019xkx}, neutrino~\cite{Dutta:2019nbn,Aguilar-Arevalo:2018wea} 
and various beam dump experiments, e.g.,~\cite{NA64:2019imj,Akesson:2018vlm}. 
There are also indirect detection proposals to probe the MeV sky by eAstrogram, GRAMS, AMEGO etc. 
The impact of MeV DM on $\Delta N_{eff}$ has been studied~\cite{Escudero:2018mvt}. 

We present here a study of MeV DM and PBH searches at the upcoming SKA radio telescope, based on the observation of photon fluxes 
generated inside a galaxy or a galaxy cluster via the IC scattering of MeV electrons-positrons on low energy CMB photons contained within that system. 
Both annihilation and decay of MeV DM are considered as sources for these MeV $e^{\pm}$.
They may also be produced in the Hawking radiation from a population of PBHs in the mass range $10^{15} - 10^{17}$ g. 
We find that, depending on the DM particle and PBH masses and the values of astrophysical parameters, 
the corresponding IC fluxes can fall (at least partially) inside the SKA frequency band.

Assuming a 100 hours of observation at the SKA and $e^+ e^-$ as the dominant final state, predicted threshold limits on 
DM annihilation rate and decay width are obtained for various DM masses in the MeV range. 
Similar limits on the PBH abundance are also presented. 
These limits are estimated using the aforementioned IC fluxes from the 
local ultra-faint galaxies Segue I and Ursa Major II, 
along with the globular cluster $\omega$-cen and the Coma cluster, for illustrative choices of 
diffusion parameters $D_0$ and $\gamma$. Because of their production mechanism,
IC fluxes and thus the SKA limits derived using them depend very weakly on the in-situ magnetic field.
By juxtaposing the SKA limits with the Planck's CMB constraints, we find that the former experiment can provide better 
probe for DM particle masses up to few tens of MeV and for PBH masses above a factor times $10^{15}$ to $10^{17}$ g 
depending on the choices of diffusion parameters in the target systems.
In parallel, we show that these SKA limits, even independently of the diffusion, 
can be stronger than the limits predicted by future MeV $\gamma$-ray 
experiments.   

We additionally demonstrate how SKA observations can be used 
to constrain the diffusion parameter space of MeV electrons inside a dwarf galaxy. 
To illustrate this, we use some example values of the DM particle mass, 
together with its annihilation and decay rates which are kept fixed at the corresponding upper limits obtained from Plank's CMB data. 
Thus, the excluded regions in the aforementioned diffusion parameter space are expected to produce detectable signals at the SKA for 
annihilation or decay scenarios that are consistent with the CMB observation.

The SKA constraints, obtained in the MeV DM or PBH parameter space for targets like dSphs, 
are estimated using some well motivated choices of the diffusion parameters. 
At the same time, the regions in the diffusion parameter space of a dSph that are possible to exclude by the future SKA observation have been identified, 
using some illustrative MeV DM scenarios consistent with existing indirect search constraints. 
The full multi-parameter space, spanned by the parameters of DM and diffusion, is difficult to constrain if one uses only the radio observation. 
However, some additional inputs from the observations of DM signals other than the radio signal may resolve the degeneracy.
For example, if the future MeV $\gamma$-ray experiments like e-ASTROGAM, AMEGO, etc., observe 
any new $\gamma$-ray signal originating from DM, then by combining such observations with the SKA radio data from dSphs, 
one can constrain both the DM and the diffusion parameters simultaneously.
In this regard, galaxy clusters such as Coma can be interesting places for the DM induced radio signal search, 
since the radio fluxes generated inside them do not depend on the diffusion.

\begin{acknowledgments}
AK thanks Tirthankar Roy Choudhury for useful discussions.
B.D., and L.S. are supported in part by the DOE Grant No. DE-SC0010813.
The work of AK was partially supported by the funding available from the Department of Atomic Energy, Government
of India, for the Regional Centre for Accelerator-based
Particle Physics (RECAPP), Harish-Chandra Research Institute.   
AK acknowledges the hospitality of the Department of Physics and Astronomy, Mitchell Institute for Fundamental Physics and Astronomy, Texas A$\&$M University, 
where the initial part of this project was formulated.
\end{acknowledgments}

\bibliographystyle{apsrev4-1}
\bibliography{biblio}

\end{document}